\documentclass[conference]{IEEEtran}
\IEEEoverridecommandlockouts
\usepackage{cite}
\usepackage{amsmath,amssymb,amsfonts,amsthm}
\usepackage{algorithm, algpseudocode}
\usepackage{graphicx}
\usepackage{textcomp}
\usepackage{xcolor}
\usepackage{tikz}
\usepackage{pgfplots}
\usetikzlibrary{external, positioning,shapes, arrows}
\usepackage{mathtools}
\usepackage{varwidth}
\usepackage[utf8]{inputenc}
\usepackage[T1]{fontenc}
\usepackage{enumitem}
\usepackage{siunitx}
\usepackage{balance}

\algnewcommand{\LeftComment}[1]{\Statex \(\triangleright\) #1}

\newtheorem{problem}{Problem}
\newtheorem{definition}{Definition}
\newtheorem{assumption}{Assumption}
\newtheorem{lemma}{Lemma}
\newtheorem*{remark}{Remark}
\def\BibTeX{{\rm B\kern-.05em{\sc i\kern-.025em b}\kern-.08em
    T\kern-.1667em\lower.7ex\hbox{E}\kern-.125emX}}

\begin{document}
	
\title{Multi-Robot Task Allocation and Scheduling \\Considering Cooperative Tasks and \\Precedence Constraints}
\author{\IEEEauthorblockN{Esther Bischoff\IEEEauthorrefmark{1}, Fabian Meyer\IEEEauthorrefmark{2}, Jairo Inga\IEEEauthorrefmark{1} and S\"oren Hohmann\IEEEauthorrefmark{1}}
	\IEEEauthorblockA{\IEEEauthorrefmark{1}Institute of Control Systems (IRS)\\
		Karlsruhe Institute of Technology (KIT)\\
		Email: esther.bischoff@kit.edu}
	\IEEEauthorblockA{\IEEEauthorrefmark{2}Research Center for Information Technology (FZI)\\
	Email: fabian.meyer@fzi.de \vspace{8pt}}
	\IEEEauthorblockA{\copyright 2020 IEEE. Personal use of this material is permitted. \\Permission from IEEE must be obtained for all other uses, in any current or future media, \\including reprinting/republishing this material for advertising or promotional purposes, \\creating new collective works, for resale or redistribution to servers or lists, \\or reuse of any copyrighted component of this work in other works.}
}
\maketitle

\begin{abstract}
In order to fully exploit the advantages inherent to cooperating heterogeneous multi-robot teams, sophisticated coordination algorithms are essential. Time-extended multi-robot task allocation approaches assign and schedule a set of tasks to a group of robots such that certain objectives are optimized and operational constraints are met. This is particularly challenging if cooperative tasks, i.e. tasks that require two or more robots to work directly together, are considered. In this paper, we present an easy-to-implement criterion to validate the feasibility, i.e. executability, of solutions to time-extended multi-robot task allocation problems with cross schedule dependencies arising from the consideration of cooperative tasks and precedence constraints. Using the introduced feasibility criterion, we propose a local improvement heuristic based on a neighborhood operator for the problem class under consideration. The initial solution is obtained by a greedy constructive heuristic. Both methods use a generalized cost structure and are therefore able to handle various objective function instances.
We evaluate the proposed approach using test scenarios of different problem sizes, all comprising the complexity aspects of the regarded problem. The simulation results illustrate the improvement potential arising from the application of the local improvement heuristic.
\end{abstract}

\section{Introduction}
In recent years the deployment of multiple robots working together towards a common goal has gained increasing attention in various application domains such as agriculture~\cite{Zhang.2017}, emergency assistance~\cite{Korsah.2011}, cleaning work~\cite{Garcia.2013} and extraterrestrial exploration~\cite{Schneider.2005}.
Multi-robot teams provide many advantages compared to single-operating robots. Tasks can be performed in parallel and the robustness of the system as a whole increases since malfunctions of single robots can possibly be compensated by the remaining robots. Furthermore, a team of heterogeneous robots can create synergies that cannot be achieved by an individual robot or even a homogeneous team. This effect is intensified if also cooperative tasks are considered, i.e. tasks which can only be performed by two or more robots working together. 
In order to fully exploit these benefits, sophisticated multi-robot task allocation (MRTA) algorithms are of great importance \cite{Gerkey.2004}, \cite[p. 1]{Korsah.2011}. Given a set of tasks to be performed by a known set of robots, these algorithms assign each task to a capable robot or a team of robots and schedule the tasks such that an executable solution results and an objective function is optimized. 
In many practical applications, this has to be done with respect to precedence constraints that exist between tasks~\cite{Zhang.2013b}, e.g. if the outcome of one task is a prerequisite for the execution of another task.

MRTA problems are often categorized using the taxonomy introduced by Gerkey and Matari\'{c} \cite{Gerkey.2004}. They differentiate for one thing between single-task (ST) and multi-task (MT) robots, dependent on whether robots can execute only one task at a time or multiple tasks simultaneously, and for another thing between single-robot (SR) and multi-robot (MR) tasks, dependent on whether tasks only require one robot for their execution or also cooperative tasks are considered. Instantaneous assignment (IA) problems are only concerned with the assignment problem whereas time-extended assignment (TA) problems additionally consider the scheduling aspect.

MRTA approaches explicitly considering multi-robot tasks have been proposed e.g. by Zhang and Parker~\cite{Zhang.2013b} who introduce a heuristic approach incorporating multi-robot tasks and precedence constraints. In \cite{Zhang.2013c}, they investigate the question of coalition formation, i.e. dynamically finding a team of robots capable of executing a specific task. Both approaches are only concerned with task allocation and do not consider task scheduling.
The time-extended problem is covered in~\cite{Zhang.2013a} where they introduce heuristics to solve the ST-MR-TA problem. The drawback of the presented approaches is that they only allow for a specific objective function and no precedence constraints are considered.

Liu and Kroll \cite{Liu.2015} introduce a memetic algorithm with a local search improvement heuristic for problems with single-robot and two-robot tasks also using a fixed objective function. They are the first to apply an improvement heuristic and additionally give executability constraints inherent to two-robot tasks. The criteria they state for detecting and repairing infeasible solutions though are limited to the two-robot task problem and do not include the consideration of precedence constraints.

Local improvement heuristics used to improve initial solutions based on various existing neighborhood definitions are a common approach in the field of vehicle routing~\cite{Ropke.2006},~\cite{Ferrucci.2013}. In this field of research, similar kinds of problems like in MRTA emerge. The problem under consideration is to route a fleet of vehicles to serve distributed customer requests such that a given objective is optimized and certain constraints are met. Also extensions to consider heterogeneous fleets of vehicles differing in velocity, capacity or the ability to serve certain types of customer requests have already been made (cf. \cite{Baldacci.2008b}, \cite{DelaCruz.2013}, \cite{Tarantilis.2007}). 
Given the similarities to the properties of MRTA problems, a direct application of the existing neighborhood operators appears conceivable. Nevertheless, this might lead to infeasible solutions arising due to a non-explicit consideration of cross-schedule dependencies. These comprise dependencies between the schedules of individual robots which influence the objective function value \cite{Korsah.2013}. They arise for example from the consideration of cooperative tasks or from waiting times due to precedence constraints.

In this paper, we propose a two-step solution approach to heterogeneous multi-robot task allocation and scheduling problems with cross schedule dependencies arising from cooperative tasks and precedence constraints. As a basis for our solution approach we introduce an easy-to-verify criterion for the feasibility, i.e. executability, of solutions to ST-MR-TA problems with cross-schedule dependencies. This allows for an adaption of the relocate neighborhood operator well known in vehicle routing \cite{Savelsbergh.1992} to make it applicable to the considered class of MRTA problems. Using the neighborhood operator, we apply an \textit{improvement heuristic} to improve the initial solution found by a \textit{constructive heuristic}. The constructive heuristic makes locally optimal choices and works similar to the MinStepSum approach presented by Zhang and Parker~\cite{Zhang.2013a}, but is enhanced to handle precedence constraints. Both heuristics use an introduced generalized objective function structure, thus being applicable to many different objective function instances.

This paper is organized as follows: We introduce the modeling and notation used within this paper and give the formal problem statement in Section \ref{sec:Notation_Problem}. In Section \ref{sec:feasibility} we introduce a feasibility definition for mission plans and state an easy-to-verify criterion for its adherence. The solution approach including a greedy constructive heuristic and an improvement heuristic for the heterogeneous multi-robot task allocation and scheduling problem with cooperative tasks and precedence constraints is given in Section \ref{sec:solution_approach}. We present simulation results in Section \ref{sec:results} and give a conclusion in Section \ref{sec:conclusion}.

\section{Modeling and Problem Formulation}\label{sec:Notation_Problem}

We first introduce the notation and modeling used throughout the paper before giving the formal problem statement.

\subsection{Notation and Modeling}\label{sec:notation}
We consider a set of tasks $T=\{t_1,\dots, t_n\}$, $n \in \mathbb{N}$, and a set of robots $R=\{r_1,\dots,r_m\}$, $m \in \mathbb{N}$. A set of robot alliances $A=\{a_1,\dots,a_k\}$, $k \in \mathbb{N}$, $k \geq m$, with ${a_j \subseteq R}$, ${\forall j \in \{1, \dots, k\}}$, specifies the possible robot coalitions.

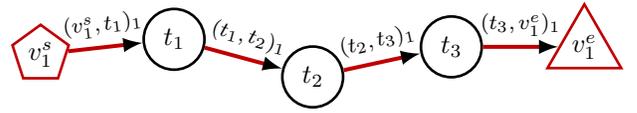
\begin{figure}[tb]
	\centering
	\usetikzlibrary{shapes.misc}
\definecolor{b1}{RGB}{91,155,213}
\definecolor{b2}{RGB}{222,235,247}

\definecolor{r1}{RGB}{192,0,0}
\definecolor{r2}{RGB}{255,205,205}

\tikzstyle{circl2} = [circle, draw, fill = white,
    text width=0.4cm, text centered]
\tikzstyle{poly5} = [regular polygon, regular polygon sides = 5, draw, fill = white, text width=0.5cm, text centered, inner sep=0pt]
\tikzstyle{poly3} = [regular polygon, regular polygon sides = 3, draw, fill = white, text width=0.4cm, text centered, inner sep=0pt]
\tikzstyle{line1} = [draw = b1, -latex]
\tikzstyle{line2} = [draw = r1, -latex]

%

\begin{tikzpicture}[node distance = 1cm, auto]
\node[poly5,text width= 4.5mm, draw= r1, line width=1](start1) {$v^s_1$};
\node[circl2, right = of start1, yshift=0.2cm,  line width = 1] (one) {$t_1$};
\node[circl2, right = of one, yshift=-0.5cm, line width = 1] (two) {$t_2$};
\node [circl2, right =  of two, yshift=0.4cm, line width = 1] (three){$t_3$};
\node[poly3, right = of three, draw=r1, line width = 1] (end1){$v^e_1$};

\draw[line2, line width = 1.5] (start1) --node[midway, sloped, above]{\footnotesize$(v^s_1,t_1)_1$} (one) ;
\draw[line2, line width = 1.5] (one) --node[midway, sloped, above]{\footnotesize$(t_1,t_2)_1$} (two);
\draw[line2, line width = 1.5] (two) --node[midway, sloped, above]{\footnotesize$(t_2,t_3)_1$} (three);
\draw[line2, line width = 1.5] (three) --node[midway, sloped, above]{\footnotesize$(t_3,v^e_1)_1$} (end1);    
\end{tikzpicture}
	\caption{Schedule of a robot represented as directed path graph. Task nodes are represent as circles, the pentagon illustrates the starting node~$v^s_1$ and the triangle the final node~$v^e_1$ of the robot~$r_1$. The arcs $(v^s_1,t_1)_1, (t_1,t_2)_1, (t_2,t_3)_1 \text{and } (t_3,v^e_1)_1 $ determine the task sequence for robot~$r_1$.}
	\label{fig:example_path_graph}
\end{figure}

For every robot $r_l \in R$ its sought schedule can be represented as a directed path graph~${G_l= \left(V_l,E_l\right)}$. The set of vertices~$V_l = \{ v^s_l, T_l, v^e_l \}$ contains all tasks~$T_l \subseteq T$ that are assigned to any alliance~$a_j$ robot~$r_l$ is part of as well as one starting node~$v^s_l$ and one end node~$v^e_l$. If a task~$t_i$ is assigned to an alliance~$a_j$ of more than one robots, it is considered as a vertex in the path graphs of all participating robots, i.e. $t_i \in V_l, \; \forall \; l: r_l \in a_j$. The nodes~$v^s_l$ and~$v^e_l$ can be related to a previously defined state or position of the robot~$r_l$ at the beginning and the end of the plan execution, respectively. Furthermore, each task~$t_i \in T$ can be associated with specific predefined properties, e.g. a position for its execution. The sequence in which robot~$r_l$ performs the assigned tasks is determined by the edges~$(v,w)_l \in E_l$ with $v, w \in V_l$. An example of a directed path graph representing the schedule of a robot~$r_1$ is presented in Fig. \ref{fig:example_path_graph}.

The overall solution, denoted as mission plan $M$, is determined by the union of the robots' individual schedules, i.e. $M=(V,E)$ with ${V=\bigcup\limits_{l \in \mathcal{M}} V_l}$, ${E=\bigcup\limits_{l \in \mathcal{M}} E_l}$, where $\mathcal{M}$ denotes the index set of the robots~$\mathcal{M} \coloneqq \{1, \dots, m\}$. We define the set~$E_{\text{in}}(v)$ to contain all incoming edges into a vertex~$v \in V$, i.e.~$E_{\text{in}}(v)$ includes all edges~$(w,v)_l$ with $w \in V$ and $l \in \mathcal{M}$.

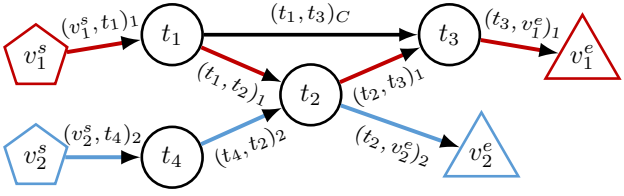
\begin{figure}[tb]
	\centering
	\usetikzlibrary{shapes.misc}
\definecolor{b1}{RGB}{91,155,213}
\definecolor{b2}{RGB}{222,235,247}

\definecolor{r1}{RGB}{192,0,0}
\definecolor{r2}{RGB}{255,205,205}

\tikzstyle{circl2} = [circle, draw, fill = white,
text width=0.4cm, text centered]
\tikzstyle{poly5} = [regular polygon, regular polygon sides = 5, draw, fill = white, text width=0.5cm, text centered, inner sep=0pt]
\tikzstyle{poly3} = [regular polygon, regular polygon sides = 3, draw, fill = white, text width=0.4cm, text centered, inner sep=0pt]
\tikzstyle{line1} = [draw = b1, -latex]
\tikzstyle{line2} = [draw = r1, -latex]

%
%

\begin{tikzpicture}[node distance = 1cm, auto]
\node[poly5, draw= r1, line width=1](start1) {$v_1^s$};
\node[circl2, right= of start1, yshift=0.3cm, line width = 1] (one) {$t_1$};
\node[circl2, right = of one, yshift=-0.8cm, line width = 1] (two) {$t_2$};
\node[circl2, right =of two, yshift=0.8cm,line width = 1] (three){$t_3$};
\node[poly3, right = of three, yshift=-0.3cm, draw=r1, line width = 1] (end1){$v_1^e$};

\draw[line2, line width = 1.5] (start1) --node[midway, sloped, above]{\footnotesize $(v_1^s,t_1)_1$} (one);
\draw[line2, line width = 1.5] (one) --node[midway, sloped, below]{\footnotesize $(t_1,t_2)_1$} (two);
\draw[line2, line width = 1.5] (two) --node[midway, sloped, below]{\footnotesize $(t_2,t_3)_1$} (three);
\draw[line2, line width = 1.5] (three) --node[midway, sloped, above]{\footnotesize$(t_3,v_1^e)_1$} (end1);

\draw[line width = 1.4,-latex] (one) --node[midway, sloped, above]{\footnotesize$(t_1,t_3)_C$} (three);

\node[poly5,below = of start1, yshift=0.5cm, draw= b1, line width=1](start2) {$v_2^s$};
\node[circl2, right = of start2, line width = 1] (four) {$t_4$};
\node[poly3, right =of two, yshift=-0.8cm, xshift=0.5cm, draw=b1, line width = 1] (end2){$v_2^e$};

\draw[line1, line width = 1.5] (start2) --node[midway, sloped, above]{\footnotesize $(v_2^s,t_4)_2$} (four);
\draw[line1, line width = 1.5] (four) --node[midway, sloped, below]{\footnotesize $(t_4,t_2)_2$} (two);
\draw[line1, line width = 1.5] (two) --node[midway, sloped, below]{\footnotesize $(t_2,v_2^e)_2$} (end2);
\end{tikzpicture}
	\caption{Example for an extended mission plan $M^+$ with four tasks, two robots and a fulfilled precedence constraint. Task $t_2$ is performed by a coalition of the robots $r_1$ and $r_2$. The black arc $(t_1,t_3)_C$ represents a precedence constraint specifying that task $t_1$ must be performed before task $t_3$.}
	\label{fig:example_mission}
\end{figure}

When generating the mission plan~$M$, precedence constraints between the tasks might have to be considered. They are specified by means of a known function ${C:T \times T \rightarrow \{0,1\}}$ with $C(t_i,t_j) = 1$ if task $t_i$ must be finished before the execution of task $t_j$, with $t_i,t_j~\in~T$, $t_i \neq t_j$, and $C(t_i,t_j) = 0$ if no such constraint exists. The precedence constraints can be included as directed edges~$(t_i,t_j)_C$ into the mission plan~$M$, if $C_{ij}=1$. We denote the mission plan extended by the set $E_C$ containing all precedence constraint arcs as $M^{+}=(V,E^+)$ with $E^+=E\cup E_C$.
In Fig.~\ref{fig:example_mission} an example of an extended mission plan with two robots, four tasks (of which one is performed by a coalition of both robots) and a precedence constraint is depicted.

Analogous to $E_{\text{in}}(v)$ we define $E_{\text{in}}^+(v)$ to be the augmented set of incoming edges to a vertex $v \in V$, additionally considering the precedence constraint edges in $E_C$, i.e. $E_{\text{in}}^+(v)$ includes $E_{\text{in}}(v)$ and all edges $(w,v)_C$ with $w \in V$.
Note that with the knowledge of $E_{\text{in}}^+(v)$ all predecessor nodes of $v$ are known.

\subsection{Generalized Cost Structure}
The objective of this paper is to optimize mission plans while being able to consider cooperative tasks and precedence constraints. Therefore, an evaluation criterion is required. We present a generic cost structure considering both costs that are static and dynamic with respect to the optimization problem. This allows for the application of the presented solution approach on a vast number of different individual objective functions which might be preferable for different problem instances.

Static costs~$c_{\text{stat}}:T\times A \rightarrow \mathbb{R}^+ \cup \{\infty\}$, ${c_{\text{stat}}= c_{\text{stat}}(t_i,a_j)}$ are associated with an alliance~$a_j \in A$ executing task~$t_i \in T$ and can be determined for every task-alliance pair prior to the optimization. For example, static costs may consider the execution duration or quality of alliance~$a_j$ performing task~$t_i$. Since the edges~$E_{\text{in}}(t_i)$ incoming to vertex~$t_i$ determine the alliance~$a_j$ assigned to task~$t_i$, the static costs can also be stated as $c_{\text{stat}}(t_i,E_{\text{in}}(t_i))$.

Cost components $c_{\text{dyn}}$ that are dynamic with respect to the optimization allow for the consideration of additional costs which might depend on the task sequence within the mission plan or on the precedence constraints. Examples for dynamic costs include moving durations and transport energy as well as idle times resulting from waiting on the coalition partners or on precedence constraints to be fulfilled. For a specific vertex~$v \in V$ the dynamic costs depend on the incoming augmented edges, i.e. $c_{\text{dyn}}: V \times E^+ \rightarrow \mathbb{R}^+$, $c_{\text{dyn}}= c_{\text{dyn}}(v,E_{\text{in}}^+(v))$. 
 
\begin{remark}
	The dynamic costs can easily be augmented to additionally take into account explicitly time dependent cost components, i.e. $c_{\text{dyn}}(v,E_{\text{in}}^+(v),\tau)$ with $\tau$ being the time. Time dependent cost components might for example arise from the consideration of time window constraints.
\end{remark}

\subsection{Problem Statement}
By means of the introduced notation, model and cost structure we are able to state the key problem of this paper:

\begin{problem}\label{problem1}
	Let the sets of robots $R$ and robot alliances $A$ and the set of vertices $V = \{v^s_1,\dots,v^s_m,t_1,\dots,t_n,v^e_1,\dots,v^e_m\}$, as well as the precedence constraint edges $E_C$ be given. We want to find directed edges $E$ such that the resulting mission plan $M=(V,E)$ is connected and feasible and an objective function $J$ dependent on the static and dynamic cost components, i.e.

	\begin{equation}\label{eq:J_general}
	\begin{aligned}
	J(M^+) = J \left( c_{\text{stat}}(t_i,E_{\text{in}}(t_i)), c_{\text{dyn}}(v,E_{\text{in}}^+(v)) \right) &\\
	 \text{for all } t_i \in T, v \in V  &
	\end{aligned}
	\end{equation}

	is minimized.
\end{problem}

As stated in Problem \ref{problem1}, only feasible mission plans are sought. We define the feasibility of mission plans in the following section.

\section{Feasibility of mission plans}\label{sec:feasibility}
The feasibility of mission plans is defined as follows:

\begin{definition}[Feasibility of a mission plan]\label{def:feasibility_plan}
	A mission plan $M$ is feasible, if it can be conducted in finite time.
	The feasibility of a mission plan comprises the following aspects:
	\begin{enumerate}[label=\textbf{D\ref{def:feasibility_plan}.\arabic*}, ref=D\ref{def:feasibility_plan}.\arabic*,leftmargin=*]
		\item \label{def:D1}
			The alliance $a_j \in A$ assigned to any task $t_i \in T$ by the mission plan $M$, must be capable of its execution.
		\item \label{def:D2}
			The mission plan $M$ must represent a topological order.
		\item \label{def:D3}
			The precedence constraints defined by $C$ 
			\begin{enumerate}[label=\textbf{\alph*)}, ref=\ref{def:D3}.\alph*,leftmargin=*]
			\item \label{def:D3a} must be consistent with one another and 
			\item must be fulfilled by the mission plan $M$. \label{def:D3b}
			\end{enumerate}
	\end{enumerate}
\end{definition}

The necessity of aspect \ref{def:D1} is obvious, since the execution time of a task will be never-ending if the alliance assigned to it is incapable of its accomplishment. We present the following assumption as an easy-to-implement method to check for the first feasibility aspect \ref{def:D1}.

\begin{assumption}\label{assumption_costs_static}
	We assume the static cost components for all $t_i \in T$, $a_j \in A$, to be of the form
	\begin{equation*}
		c_{\text{stat}}(t_i,a_j): T \times A \rightarrow
		\begin{cases}
		\infty & \text{\parbox[c]{0.5\linewidth}{if alliance $a_j$ is incapable to execute task $t_i$,}}\\
		\mathbb{R}^+ & \text{else.}
		\end{cases}
	\end{equation*}
\end{assumption}

We furthermore present Lemma \ref{lemma:acyclic_constraints} as an important insight to examine the aspects \ref{def:D2} and \ref{def:D3} of Definition~\ref{def:feasibility_plan} necessary for the feasibility of a mission plan.

\begin{lemma}\label{lemma:acyclic_constraints}
	If the mission plan $M$ is feasible, then the directed graph $M^{+}$ of the feasible mission plan $M$ extended by the precedence constraint edges $E_C$ is acyclic.
\end{lemma}

\begin{IEEEproof}
A topological ordering of a directed graph is possible if and only if the graph is acyclic (cf.~\cite[Ch.~4.2]{Sedgewick.2012}). Therefore, \ref{def:D2} holds if and only if $M$ is acyclic. For the same reason, the precedence constraints fulfill \ref{def:D3a} if and only if the graph~$G_C=(V,E_C)$ only containing the precedence constraint edges is acyclic. When adding the precedence constraint edges $E_C$ to $M$, which results in the augmented mission graph $M^+$, two cases have to be considered regarding the feasibility of $M$:
\begin{itemize}
	\item The sets of robots assigned to the tasks $t_i, t_j \in T$ which are related by a precedence constraint edge~${(t_i,t_j)_C \in E_C}$ are disjoint. In this case, the edge~$(t_i,t_j)_C$ does not add a cycle to the acyclic graph $M$ and the precedence constraint defined by $(t_i,t_j)_C$ can always be fulfilled if the alliance assigned to task~$t_j$ ensures to wait with its execution until task~$t_i$ is finished. Capable alliance-task-assignments (\ref{def:D1}) ensure the potential time increment to be bounded.	
	\item The robot alliances assigned to the tasks~${t_i, t_j \in T}$ which are related by a precedence constraint edge~${(t_i,t_j)_C \in E_C}$ are not disjoint. 
	The edge~$(t_i,t_j)_C$ only closes a cycle in $M^+$, if at least one robot $r_l \in R$ assigned to both tasks $t_i$ and $t_j$ violates the precedence constraint, which means that \ref{def:D3b} would not be fulfilled and $M$ would be infeasible.
\end{itemize}
\end{IEEEproof}

Using the results of Lemma \ref{lemma:acyclic_constraints} and combining it with Assumption \ref{assumption_costs_static}, Lemma \ref{lemma:feasibility_total} gives necessary and sufficient conditions for the feasibility of mission plans according to Definition~\ref{def:feasibility_plan}.

\begin{lemma}[Feasibility of a mission plan]\label{lemma:feasibility_total}
	Let Assumption \ref{assumption_costs_static} hold. Then, a mission plan $M$ is feasible w.r.t. Definition \ref{def:feasibility_plan} if and only if
	\begin{enumerate}[label=\textbf{L\ref{lemma:feasibility_total}.\arabic*}, ref=L\ref{lemma:feasibility_total}.\arabic*,leftmargin=*]
	\item \label{cstat_finite} 
		the static costs of all task vertexes $t_i \in T$ are finite, i.e.
		\begin{equation*}
		c_{\text{stat}}(t_i, E_{\text{in}}(t_i)) < \infty, \hspace{0.5cm}\forall t_i \in T,
		\end{equation*}  
	\item \label{Mplus_acyclic}
		and the directed graph of the augmented mission plan $M^+$ is acyclic.
\end{enumerate}
\end{lemma}

\begin{IEEEproof}
With the results of Lemma~\ref{lemma:acyclic_constraints}, \ref{Mplus_acyclic} gives a necessary and sufficient condition for the feasibility aspects \ref{def:D2} and \ref{def:D3} to be fulfilled. Since Assumption~\ref{assumption_costs_static} holds, \ref{cstat_finite} gives necessary and sufficient condition to ensure that $M$ fulfills~\ref{def:D1}.
\end{IEEEproof}

\section{Solution Approach}\label{sec:solution_approach}

Before presenting our two-step solution approach comprising a constructive and an improvement heuristic in detail, we assume the following assumptions to hold:

\begin{assumption}\label{assumption_solution_existance}
	For a given instance of Problem~\ref{problem1} for every task $t_i \in T$ at least one capable alliance $a_j \in A$ with $c_{\text{stat}}(t_i,a_j) \in \mathbb{R}^+$ exists.
\end{assumption}

\begin{assumption}\label{assumption_solution_existance2}
	For a given instance of Problem~\ref{problem1} the graph~$G_C=(V,E_C)$ is acyclic.
\end{assumption}

Assumption \ref{assumption_solution_existance2} ensures \ref{def:D3a} to be fulfilled by the a priori given precedence constraints. Thus, Assumptions \ref{assumption_solution_existance} and \ref{assumption_solution_existance2} are made to assure meaningful problem instances.

\subsection{Constructive Heuristic}

The constructive heuristic iteratively calculates the effect every new assignment would have on the objective function and chooses the one that increases the objective function the least until all tasks~$t_i \in T$ have been assigned. The idea is similar to the MinStepSum algorithm introduced by Zhang and Parker \cite{Zhang.2013a}, but we expand it to handle the generalized objective function given by \eqref{eq:J_general}. Furthermore we augment the method to additionally allow for the direct consideration of precedence constraints. To do so, we split the tasks into the sets of \textit{executable tasks}~$\Lambda$ and \textit{non executable tasks}~$\overline{\Lambda}$ with $\Lambda \cap \overline{\Lambda} = \emptyset$. The set~$\overline{\Lambda}$ contains all tasks with a nonempty set of unassigned precedence tasks. The two sets are initialized with elements $t_i$, where

\begin{equation}\label{eq:init_lambda}
t_i \in
\begin{cases}
\Lambda & \text{if }C(t,t_i)=0 \; \forall \; t \in T \\
\overline{\Lambda} & \text{else}.
\end{cases}
\end{equation}

\begin{algorithm}[tb]
	\caption{Constructive Heuristic}\label{alg:construction}
	\begin{algorithmic}[1]
		\Require $R, A, \Lambda, \overline{\Lambda}, V\backslash T, E_C$
		\ForAll{$r_l \in R$} \Comment{Initialization: path graphs of robots}\label{alg:init_start}
			\State $E_l \gets \emptyset$, $V_l \gets \{v_l^s\}$, $G_l=(V_l,E_l)$
		\EndFor 
		\State{$M \gets \cup_{l \in \mathcal{M}} G_l$} \Comment{Initialize: mission plan}
		\State{$J \gets J(M^+)$}\label{alg:init_end} \Comment{Initialize: objective function}
		\While{$\Lambda \neq \emptyset$}
			\State $\Delta_{min} \leftarrow \infty$ \Comment{Initialize: objective function increment}
			\Statex \Comment{\parbox[t]{0.9\linewidth}{For every executable tasks-alliance pair calculate cost increment of the assignment:}}
			\ForAll{$t_i \in \Lambda$}\label{alg:local_choice_start}
				\ForAll{$a_j \in A$}
					\ForAll{$r_l \in a_j$} 
						\State $v_{\text{leaf}} \gets \{v \in V_l: \nexists w\in V_l: (v,w)_l \in E_l\}$
						\State $\tilde{V}_l \gets \{V_l, t_i\}$, $\tilde{E}_l \gets \{E_l, (v_{\text{leaf}},t_i)_l\}$ 
						\State $\tilde{G}_l=(\tilde{V}_l,\tilde{E}_l)$
					\EndFor 
					\State{$\tilde{M} \gets \cup_{l \in \mathcal{M}} \tilde{G}_l$} 
					\State{$J_{\Delta}= J(\tilde{M}^+)-J$}\label{alg:cost_incr}\label{alg:local_choice_end1}
					\If{$J_{\Delta}\leq \Delta_{min}$} \label{alg:local_choice_best_start}
						\State $M_{\text{min}} \gets \tilde{M}$ \Comment{Remember best assignment}
						\State{$\Delta_{min} \gets J_{\Delta}$} \Comment{\parbox[t]{.46\linewidth}{Remember smallest objective function increment}\vspace{0.15cm}}
						\State{$t_{min} \gets t_i$} \Comment{\parbox[t]{.46\linewidth}{Remember assigned task}}
					\EndIf 
				\EndFor
			\EndFor
			\State $M \gets M_{\text{min}}$
			\State $J \gets J + \Delta_{min}$ \label{alg:local_choice_end2}
			\Statex \Comment{\parbox[t]{0.9\linewidth}{Check if~$t_{\text{min}}$ was the only remaining precedence constraint to any~$t_i \in \overline{\Lambda}$:}}
			\ForAll {$t_i \in \overline{\Lambda}$ with $C(t_{\text{min}},t_i)=1$} \label{alg:executable_check_start}
				\If{$ C(t,t_i)=0 \; \forall t \in \{\Lambda \cup \overline{\Lambda}\}\backslash\{t_{\text{min}}\}$}
					\State{$\Lambda \leftarrow \{\Lambda, t_i\}$} \Comment{Add $t_i$ to $\Lambda$}
					\State{$\overline{\Lambda} \gets \overline{\Lambda} \backslash \{t_i\}$} \Comment{Delete $t_i$ from $\overline{\Lambda}$}
				\EndIf
			\EndFor \label{alg:executable_check_end}
			\State{$\Lambda \gets \Lambda \backslash \{t_{\text{min}}\}$} \Comment{Delete $t_{\text{min}}$ from $\Lambda$}
		\EndWhile
		\ForAll{$r_l \in R$}\label{alg:final_node_start} 
			\State $v_{\text{leaf}} \gets \{v \in V_l: \nexists w\in V_l: (v,w)_l \in 	E_l\}$
			\State $V_l \gets \{V_l, v_l^e\}$, $\tilde{E}_l \gets \{E_l, (v_{\text{leaf}},v_l^e)_l\}$ 
			\State $G_l=(V_l,E_l)$
		\EndFor
		\State{$M_{\text{init}} \gets \cup_{l \in \mathcal{M}} G_l$}
		\State{$J_{\text{init}} \gets J(M_{\text{init}}^+)$}\label{alg:final_node_end}\\
		\Return{$M_{\text{init}}$, $J_{\text{init}}$}
	\end{algorithmic}
\end{algorithm}

The detailed constructive heuristic is given in Algorithm~\ref{alg:construction} and works as follows: The algorithm is given the sets $R$, $A$, and $V$ (divided into the sets $V\backslash T$ and $\Lambda$, $\overline{\Lambda}$ which are initialized according to \eqref{eq:init_lambda} such that $\Lambda \cup \overline{\Lambda}=T$) as well as the precedence constraints $E_C$. For every robot~$r_l \in R$, the path graph~$G_l$ is initialized with an empty graph containing only the robots initial vertex $v_l^s$. Using these initial path graphs, the mission plan is initialized and the initial objective function is calculated~(lines~\algref{alg:construction}{alg:init_start} to~\algref{alg:construction}{alg:init_end}). For every not yet assigned executable task~$t_i \in \Lambda$ and every robot alliance~$a_j \in A$, the increment of the objective function resulting from the respective assignment is calculated by adding $t_i$ as leaf to the path graphs of the respective robots and calculating the objective function value increment for the resulting intermediate augmented mission plan $\tilde{M}$ (lines~\algref{alg:construction}{alg:local_choice_start}~to~\algref{alg:construction}{alg:local_choice_end1}). Out of all possible assignments the one with the smallest objective function increment is chosen~(lines~\algref{alg:construction}{alg:local_choice_best_start} to~\algref{alg:construction}{alg:local_choice_end2}) . The assigned task is deleted from the set of executable tasks~$\Lambda$ and all tasks from the set of non executable tasks~$\overline{\Lambda}$ that became executable with the most recent assignment are transferred to the set of executable tasks~$\Lambda$ (lines~\algref{alg:construction}{alg:executable_check_start}~to~\algref{alg:construction}{alg:executable_check_end}). The procedure repeats until all tasks have been assigned. The algorithm terminates by adding the robots final nodes $v_l^e$, $\forall r_l \in R$, to the respective robots' path graphs and determining the resulting mission plan $M_{\text{init}}$ and the respective objective function value $J_{\text{init}}$ (lines~\algref{alg:construction}{alg:final_node_start}~to~\algref{alg:construction}{alg:final_node_end}).

\begin{remark}
	Assumption~\ref{assumption_solution_existance} implies that for every task
	$t_i \in T$ an assignment will be found for which \ref{def:D1} is fulfilled. Furthermore, the explicit consideration of precedence constraints by means of the sets~$\Lambda$ and~$\overline{\Lambda}$ guarantees their adherence according to \ref{def:D3b} and Assumption~\ref{assumption_solution_existance2} ensures \ref{def:D3a} such that also \ref{def:D3} will be fulfilled by the mission plan resulting from the constructive heuristic. The adherence of \ref{def:D2} is ensured by the fact that all newly assigned tasks are added as leafs to the path graphs of the respective robots which means that tasks assigned to coalitions of several robots are assured to have the same sequence within the individual path graphs of the robots. Therefore the solution found by the constructive heuristic given in Algorithm~\ref{alg:construction} will always be feasible w.r.t Definition \ref{def:feasibility_plan}.
\end{remark}

\subsection{Improvement Heuristic}
We apply a local search to further improve the initial mission plan~$M_{\text{init}}$ found by the constructive heuristic. In every iteration the currently best solution is modified using a neighborhood operator which is based on the relocate neighborhood first introduced by Savelsbergh and Goetschalckx \cite{Savelsbergh.1992} for the routing problem of a homegeneous fleet of vehicles. The neighborhood operator given in Definition~\ref{def:neighborhood} expands the original relocate operation to be applicable to mission plans~$M$ for heterogeneous robotic teams with cooperative tasks and precedence constraints.

\begin{definition}[\textbf{neighborhood of a mission plan}]\label{def:neighborhood}
	The neighborhood of a mission plan~$M$ contains all feasible mission plans $\tilde{M}$ that result from relocating one task~$t_i \in T$ out of the path graphs of the alliance~$a_j$ it is assigned to by~$M$, to any position of the path graphs of the robots of any capable alliance~$a_{\tilde{l}} \in A$.
\end{definition}

Using the neighborhood of Definition~\ref{def:neighborhood} and the results of Lemma~\ref{lemma:feasibility_total}, the local search improvement heuristic is given in Algorithm~\ref{alg:improvement}.
\begin{algorithm}[tb]
	\caption{Improvement Heuristic}\label{alg:improvement}
	\begin{algorithmic}[1]
		\Require~$M_{\text{init}}, J_{\text{init}}, R, A, C$
		\LeftComment{Initialization:}
		\State{$J_{\text{best}} \gets J_{\text{init}}$} \label{alg:impr_init1}
		\State{$M_{\text{best}} \gets M_{\text{init}}$} \label{alg:impr_init2}
		\While{Stopping criterion not fulfilled}
		\ForAll{$t_i \in M_{\text{best}}$} \label{alg:impr_reassign_start}
		\ForAll{$a_{\tilde{l}} \in A$}
		\If{$c_{\text{stat}}(t_i, a_{\tilde{l}}) < \infty$}
		\State{\parbox[t]{.7\linewidth}{Determine all possible reassignments of~$t_i$ to the path graphs~$G_{\tilde{l}}: r_{\tilde{l}} \in a_{\tilde{l}}$}} \label{alg:impr_reassign_end}
		\ForAll{possible reassignments $\tilde{M}$}
		\State{\parbox[t]{.65\linewidth}{Check whether the resulting augmented mission plan $\tilde{M}^+$ is acyclic}} \label{alg:impr_cycle_check}
		\If{$\tilde{M}^+$ is acyclic} \label{alg:impr_comparisson_start}
		\State{Calculate $J(\tilde{M}^+)$}
		\If{$J(\tilde{M}^+)<J_{\text{best}}$}
		\State{$M_{\text{best}} \gets \tilde{M}$}
		\State{$J_{\text{best}} \gets J(\tilde{M}^+)$}
		\EndIf
		\EndIf \label{alg:impr_comparisson_end}
		\EndFor
		\EndIf
		\EndFor
		\EndFor
		\EndWhile \\
		\Return{$M_{\text{best}}, J_{\text{best}}$}
	\end{algorithmic}
\end{algorithm}
The mission plan and the objective function are initialized with the results of the constructive heuristic (lines~\algref{alg:improvement}{alg:impr_init1}~to~\algref{alg:improvement}{alg:impr_init2}). In every iteration, the neighborhood of the currently best mission plan is determined and evaluated and the best neighboring mission plan is chosen. To determine and evaluate the neighborhood, for all tasks~$t_i \in T$, all possible reassignment positions within the path graphs of the robots of every alliance~$a_{\tilde{l}} \in A$ that is capable of the execution of task~$t_i$ (i.e. the static costs $c_{\text{stat}}(t_i,a_{\tilde{l}})$ are bounded) are determined (lines~\algref{alg:improvement}{alg:impr_reassign_start}~to~\algref{alg:improvement}{alg:impr_reassign_end}). To assure the feasibility of the resulting new mission plan~$\tilde{M}$, Lemma~\ref{lemma:acyclic_constraints} is applied and it is determined whether the augmented mission plan~$\tilde{M}^+$ contains cycles  (line~\algref{alg:improvement}{alg:impr_cycle_check}). If~$\tilde{M}^+$ is acyclic and~$\tilde{M}$ therefore a feasible neighboring mission plan, its objective function value is determined and assessed in comparison to the currently best plan found (lines~\algref{alg:improvement}{alg:impr_comparisson_start}~to~\algref{alg:improvement}{alg:impr_comparisson_end}).
This procedure repeats until a stopping criterion is fulfilled, e.g. the improvement in the objective function~$J$ between to iterations falls below a previously determined threshold or a previously determined maximum number of iterations is reached.

To conduct the acyclicity check (line~\algref{alg:improvement}{alg:impr_cycle_check}), any cycle search for digraphs can be applied. Since only the statement about acyclicity and not the potentially existing cycles are of interest, we implemented an algorithm based on iteratively removing leafs from the augmented mission plan $M^+$, similar to the algorithm of Kahn \cite{Kahn.1962}.

\section{Simulation Results}\label{sec:results}
The experimental setup and the structure of the objective function used to evaluate the introduced MRTA approach are described in the following section followed by the presentation and discussion of the simulation results.

\subsection{Experimental Setup}\label{sec:exp_setup}
To evaluate the proposed solution approach we set up a generalized simulation framework for MRTA problems with precedence constraints and cooperative tasks. It consists of four different types of tasks and three different mobile robots. Each task type can be processed by a subset of the considered alliances $A = \{\{r_1\},\{ r_2\}, \{r_3\}, \{r_1,r_2\}, \{r_1,r_3\}, \{r_2,r_3\}\}$.

Based on the different task types we define six different problem classes:  $3\mathcal{A}1\mathcal{B}\mathcal{C}\mathcal{D}$, $3\mathcal{A}2\mathcal{B}\mathcal{C}\mathcal{D}$, $3\mathcal{A}3\mathcal{B}\mathcal{C}\mathcal{D}$, $6\mathcal{A}1\mathcal{B}\mathcal{C}\mathcal{D}$, $6\mathcal{A}3\mathcal{B}\mathcal{C}\mathcal{D}$ and $6\mathcal{A}3\mathcal{B}\mathcal{C}\mathcal{D}$.
The first number in these coded problem classes describes the number of tasks of type $\mathcal{A}$ whereas the second number denotes the number of tasks of each type $\mathcal{B}$, $\mathcal{C}$ and $\mathcal{D}$. To indicate the type an individual task belongs to, we add a superscript to the tasks $t_i$. The index $i$ starting at $1$ in ascending order is first assigned to all task of type $\mathcal{A}$ followed be the tasks of type $\mathcal{B}$, $\mathcal{C}$ and $\mathcal{D}$. This results for example in problem class $3\mathcal{A}2\mathcal{B}\mathcal{C}\mathcal{D}$ having the set of tasks $T_{3\mathcal{A}2\mathcal{B}\mathcal{C}\mathcal{D}}=\{t_1^{\mathcal{A}},t_2^{\mathcal{A}},t_3^{\mathcal{A}}, t_4^{\mathcal{B}}, t_5^{\mathcal{B}}, t_6^{\mathcal{C}}, t_7^{\mathcal{C}}, t_8^{\mathcal{D}},t_9^{\mathcal{D}}\}$.

Let $\vert\mathcal{A}\vert$, $\vert\mathcal{B}\vert$ and $\vert\mathcal{C}\vert$ denote the number of tasks of type $\mathcal{A}$, $\mathcal{B}$ and $\mathcal{C}$, respectively, within a given problem class. For all problem classes we assume the following types of precedence constraints to exist exactly once.

\begin{itemize}
	\item $(t^\mathcal{A}_1, t^\mathcal{A}_2)_C$: The first type $\mathcal{A}$ task must be completed before processing of the second type $\mathcal{A}$ task can begin.
	\item  $(t^\mathcal{A}_3, t^\mathcal{B}_{\vert\mathcal{A}\vert+1})_C$: The third type $\mathcal{A}$ task must be completed before processing of the first type $\mathcal{B}$ task can begin.
	\item  $(t^\mathcal{C}_{\vert\mathcal{A}\vert+\vert\mathcal{B}\vert+1}, t^\mathcal{D}_{\vert\mathcal{A}\vert+\vert\mathcal{B}\vert+\vert\mathcal{C}\vert+1})_C$: The first type $\mathcal{C}$ task must be completed before processing of the first type $\mathcal{D}$ task can begin.
\end{itemize}

To generate specific problem instances for each problem class, we consider every task to be associated with a position for its execution.
The individual task positions depend on the task indices $i$ and are located in the Cartesian plane at
\begin{align}
	x(t_i) &= L_0 \cos(\theta_0(t_i)) + L_1 \cos(\theta_1)  \\
	y(t_i) &= L_0 \sin(\theta_0(t_i)) + L_1 \sin(\theta_1)
\end{align}
with
\begin{alignat}{3}
L_0 &= 50m, && & L_1 & \in [0m, 10m], \nonumber \\
\theta_0(t_i) &= 2 \pi \frac{i}{|T|} + \frac{\pi}{|T|}, \quad && & \theta_1 &\in [0, 2\pi].\nonumber
\end{alignat}

The values for $L_1$ and $\theta_1$ are equally distributed with respect to the given intervals over all problem instances of a certain problem class.
Thus, for each task of a problem class an area is defined in which the corresponding task must be located. By these definitions we aim to ensure a high comparability of all the problem instances of a given problem class. For all robots $r_l$, $l \in \{1,2,3\}$, their starting node $v^s_l$ is associated with the initial position $(0,0)$ at the origin of the coordinate system whereas the position of their end node $v^e_l$ is set to be arbitrary.

\subsection{Structure of the Objective Function}\label{sec:res_objfct}
For every task-alliance pair the static cost component $c_{\text{stat}}(t_i, a_j)$ represents the duration needed by alliance $a_j$, $j \in \{1,\dots,6\}$ to perform task $t_i \in T$. The respective values dependent on the task types are given in Table \ref{tab::static_duration}.

\begin{table}[tb]
	\centering
	\caption{Task duration in (s) for every task-alliance pair}
	\begin{tabular}{ccccc}
		\hline
		Alliance & Type $\mathcal{A}$ & Type $\mathcal{B}$ & Type $\mathcal{C}$ & Type $\mathcal{D}$ \\
		\hline
		$\{r_1\}$&100 & $\infty$ & $\infty$ & $\infty$ \\
		$\{r_2\}$&100 & $\infty$ & $\infty$ & $\infty$ \\
		$\{r_3\}$&100 & $\infty$ & $\infty$ & 200 \\
		$\{r_1, r_2\}$ & $\infty$ & 110 & $\infty$ & $\infty$ \\
		$\{r_1, r_3\}$& $\infty$ & 100 & 100 & $\infty$ \\
		$\{r_2, r_3\}$& $\infty$ & $\infty$ & $\infty$ & 100 \\
		\hline
	\end{tabular}
	\vspace{0.3cm}
	\label{tab::static_duration}
\end{table}

The dynamic cost components include:
\begin{itemize}
	\item
		$c_{\text{dyn}}^1(v,E_{\text{in}}^+(v))$: For every edge $(w,v)_l \in E_{\text{in}}^+(v)$ we calculate the driving time $\tau_{d,v}(r_l)$ needed by robot $r_l$ to travel from the position of its previous node $w$ to the position of $v$.
		To calculate the individual traveling times we use the euclidean distance between the positions of $w$ and $v$ and the robot's individual velocities $v(r_l)$, which are set to be $v(r_1) = v(r_2) = \SI{2}{\metre\per\second}$ and $v(r_3) = \SI{1}{\metre\per\second}$.
	\item
		$c_{\text{dyn}}^2(t_i,E_{\text{in}}^+(t_i))$: For every edge $(w,t_i)_l \in E_{\text{in}}^+(t_i)$ the waiting time $\tau_{w,t_i}(r_l)$ of robot $r_l$ resulting from waiting on coalition partners to reach the position of $t_i$ or on precedence constraints for $t_i$ to be fulfilled is determined.
	\item
		$c_{\text{dyn}}^3(v,E_{\text{in}}^+(v))$: For every edge $(w,v)_l \in E_{\text{in}}^+(v)$ we calculate the euclidean distance $d_{v}(r_l)$ driven by robot~$r_l$ to travel from the position of its previous node $w$ to the position of $v$.
\end{itemize}

In the objective function we consider the total mission duration given by latest finishing time over all robots $r_l \in R$, i.e.
\begin{equation}
	\begin{aligned}
		J_1(M^+)= \max_{l \in \{1,2,3\}} &\left\{ \sum_{t_i \in V_l} \left( c_{\text{stat}}(t_i, E_{\text{in}}(t_i)) +\tau_{w,t_i}(r_l) \right) \right.  \\
		& \left. + \sum_{v \in V_l} \tau_{d,v}(r_l) \right\},
	\end{aligned}
\end{equation}
as well as the average finishing time of the robots
\begin{equation}
\begin{aligned}
	J_2(M^+)= \frac{1}{3} \sum_{l=1}^{3} &\left\{ \sum_{t_i \in V_l} \left( c_{\text{stat}}(t_i, E_{\text{in}}(t_i)) +\tau_{w,t_i}(r_l) \right) \right. + \\
	& \left. \sum_{v \in V_l} \tau_{d,v}(r_l) \right\}
\end{aligned}
\end{equation}
and the sum over all driven distances divided by the number of robots
\begin{equation}
J_3(M^+)= \frac{1}{3} \sum_{l=1}^{3} \sum_{v \in V_l} d_{v}(r_l) .
\end{equation}
This choice of weighting factors reflects the fact that in many practical relevant scenarios the minimization of the total mission duration comes with the highest priority.

To test and validate our approach we evaluated 100 problem instances for each problem class on an Intel(R) Core(TM) i5-8250U CPU at 1.6GHz and 8GB RAM with a Windows 10 operating system. Our optimization method was implemented using MATLAB R2017b.

\subsection{Results}
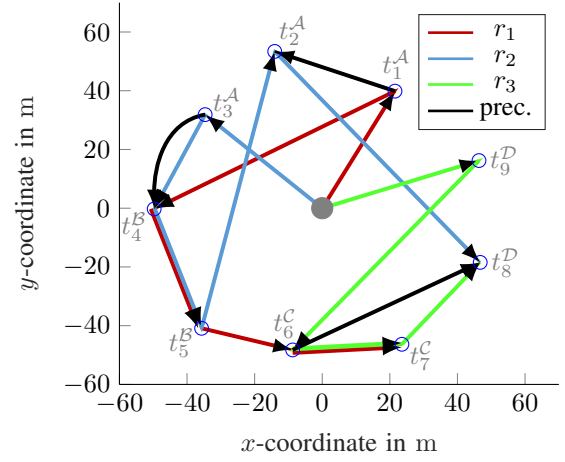
\begin{figure}[tb]
	\centering
%
%
\definecolor{mycolor1}{rgb}{0.00000,0.44700,0.74100}%
\definecolor{mycolor2}{rgb}{0.85000,0.32500,0.09800}%
\definecolor{mycolor3}{rgb}{0.92900,0.69400,0.12500}%
\definecolor{mycolor4}{RGB}{91,155,213}%
\definecolor{mycolor5}{rgb}{0.49400,0.18400,0.55600}%
\definecolor{red}{RGB}{192,0,0}
\definecolor{green}{RGB}{83,255,47}
\tikzstyle{line1} = [line width = 1.5, draw = red, -latex]
\tikzstyle{line2} = [line width = 1.5, draw = mycolor4, -latex]
\tikzstyle{line3} = [line width = 1.5, draw = green, -latex]
\tikzstyle{line4} = [line width = 1.5, draw = black, -latex]
\begin{tikzpicture}

\begin{axis}[%
width=2.3in,
height=2in,
at={(0.758in,0.481in)},
scale only axis,
clip=false,
xmin=-60,
xmax=70.01,
xlabel style={font=\color{white!15!black}},
xlabel={$x$-coordinate in $\si{\metre}$},
ymin=-60,
ymax=70,
ylabel style={font=\color{white!15!black}},
ylabel={$y$-coordinate in $\si{\metre}$},
axis background/.style={fill=white},
title style={font=\bfseries},
axis x line*=bottom,
axis y line*=left,
legend style={at={(0.5, -0.0)}, anchor = north east, draw=white!15!black},
legend pos = north east
]
\addplot [color=blue, draw=none, mark=o, mark options={solid, blue, scale = 1.3}, forget plot]
  table[row sep=crcr]{%
21.5226153329081	39.8647172246807\\
};


\node[right, align=left, font=\color{gray}]
at (axis cs:15.094,48.865) (A1) {$t^{\mathcal{A}}_1$};
\addplot [color=blue, draw=none, mark=o, mark options={solid, blue, scale = 1.3}, forget plot]
  table[row sep=crcr]{%
-14.0620789873635	53.3854639899131\\
};
\node[right, align=left, font=\color{gray}]
at (axis cs:-15.491,60.385) {$t^{\mathcal{A}}_2$};
\addplot [color=blue, draw=none, mark=o, mark options={solid, blue, scale = 1.3}, forget plot]
  table[row sep=crcr]{%
-34.6383262329816	31.8621878887854\\
};
\node[right, align=left, font=\color{gray}]
at (axis cs:-35.067,35.862) {$t^{\mathcal{A}}_3$};
\addplot [color=blue, draw=none, mark=o, mark options={solid, blue, scale = 1.3}, forget plot]
  table[row sep=crcr]{%
-49.7166532901225	-0.249272630880901\\
};
\node[right, align=left, font=\color{gray}]
at (axis cs:-62.145,-6.249) {$t^{\mathcal{B}}_4$};
\addplot [color=blue, draw=none, mark=o, mark options={solid, blue, scale = 1.3}, forget plot]
  table[row sep=crcr]{%
-35.6890668539422	-40.8904205551716\\
};
\node[right, align=left, font=\color{gray}]
at (axis cs:-48.118,-45.89) {$t^{\mathcal{B}}_5$};
\addplot [color=blue, draw=none, mark=o, mark options={solid, blue, scale = 1.3}, forget plot]
  table[row sep=crcr]{%
-8.75596898205058	-48.2560095181754\\
};
\node[right, align=left, font=\color{gray}]
at (axis cs:-17.185,-39.256) {$t^{\mathcal{C}}_6$};
\addplot [color=blue, draw=none, mark=o, mark options={solid, blue, scale = 1.3}, forget plot]
  table[row sep=crcr]{%
23.5669353153818	-46.3332244469052\\
};
\node[right, align=left, font=\color{gray}]
at (axis cs:23.138,-50.333) {$t^{\mathcal{C}}_7$};
\addplot [color=blue, draw=none, mark=o, mark options={solid, blue, scale = 1.3}, forget plot]
  table[row sep=crcr]{%
46.7391087346741	-18.4442847735162\\
};
\node[right, align=left, font=\color{gray}]
at (axis cs:47.311,-18.444) {$t^{\mathcal{D}}_8$};
\addplot [color=blue, draw=none, mark=o, mark options={solid, blue, scale = 1.3}, forget plot]
  table[row sep=crcr]{%
46.377664375942	16.2227128084728\\
};
\node[right, align=left, font=\color{gray}]
at (axis cs:46.949,16.223) {$t^{\mathcal{D}}_9$};
\addplot [color=gray, draw=none, mark size=4.0pt, mark=*, mark options={solid, fill=gray, gray}, forget plot]
  table[row sep=crcr]{%
0	0\\
};
\addplot [color=red, line width=1.0pt]
  table[row sep=crcr]{%
0	0\\
21.5226153329081	39.8647172246807\\
-49.7166532901225	-0.249272630880901\\
-35.6890668539422	-40.8904205551716\\
-8.75596898205058	-48.2560095181754\\
23.5669353153818	-46.3332244469052\\
};
\draw[line1](axis cs:0,0) -- (axis cs:21.5226153329081,	39.8647172246807);
\draw[line1](axis cs:21.5226153329081,	39.8647172246807) -- (axis cs:-49.7166532901225,	-0.249272630880901);
\draw[line1](axis cs:-50.7166532901225,	-0.249272630880901) -- (axis cs:-36.6890668539422,	-40.8904205551716);
\draw[line1](axis cs:-35.6890668539422,	-40.8904205551716) -- (axis cs:-8.75596898205058,	-48.2560095181754);
\draw[line1](axis cs:-8.75596898205058,	-49.2560095181754) -- (axis cs:23.5669353153818,	-47.3332244469052);
\addlegendentry{$r_1$}

\addplot [color=mycolor4, line width=1.0pt]
  table[row sep=crcr]{%
0	0\\
-34.6383262329816	31.8621878887854\\
-49.7166532901225	-0.249272630880901\\
-35.6890668539422	-40.8904205551716\\
-14.0620789873635	53.3854639899131\\
46.7391087346741	-18.4442847735162\\
};
\draw[line2](axis cs:0,0) -- (axis cs:-34.6383262329816,	31.8621878887854);
\draw[line2](axis cs:-34.6383262329816,	31.8621878887854) -- (axis cs:-49.7166532901225,	-0.249272630880901);
\draw[line2](axis cs:-49.5166532901225,	-0.249272630880901) -- (axis cs:-35.6890668539422,	-40.3904205551716);
\draw[line2](axis cs:-35.6890668539422,	-40.8904205551716) -- (axis cs:-14.0620789873635,	53.3854639899131);
\draw[line2](axis cs:-14.0620789873635,	53.3854639899131) -- (axis cs:46.7391087346741,	-18.4442847735162);
\addlegendentry{$r_2$}

\addplot [color=green, line width=1.0pt]
  table[row sep=crcr]{%
0	0\\
46.377664375942	16.2227128084728\\
-8.75596898205058	-48.2560095181754\\
23.5669353153818	-46.3332244469052\\
46.7391087346741	-18.4442847735162\\
};
\draw[line3](axis cs:0,0) -- (axis cs:46.377664375942,	16.2227128084728);
\draw[line3](axis cs:46.377664375942,	16.2227128084728) -- (axis cs:-8.75596898205058,	-48.2560095181754);
\draw[line3](axis cs:-8.75596898205058,	-47.7560095181754) -- (axis cs:23.5669353153818,	-45.8332244469052);
\draw[line3](axis cs:23.5669353153818,	-46.3332244469052) -- (axis cs:46.7391087346741,	-18.4442847735162);
\addlegendentry{$r_3$}
\draw[line4](axis cs:21.5226153329081,	39.8647172246807) -- (axis cs:-14.0620789873635,	53.3854639899131);
\draw[line4](axis cs:-34.6383262329816,	31.8621878887854) to [out = 190, in = 90]  (axis cs:-49.7166532901225,	-0.249272630880901);
\draw[line4](axis cs:-8.75596898205058,	-48.2560095181754) -- (axis cs:46.7391087346741,	-18.4442847735162);

\addplot[color=black, line width=1.0pt]
coordinates {%
	(21.5226153329081,	39.8647172246807)
   (-14.0620789873635,	53.3854639899131)
};

\addlegendentry{prec.}

\end{axis}
\end{tikzpicture}%
	\caption{Resulting mission graph representing a locally optimal solution of an instance of problem class $3\mathcal{A}2\mathcal{B}\mathcal{C}\mathcal{D}$. The colored paths are associated to a robots individual schedule. Black edges illustrate precedence constraints.}
	\label{fig:paths}
\end{figure}

\begin{figure}[tb]
	\centering
%
%
\definecolor{mycolor1}{rgb}{0.00000,1.00000,0.66667}%
\definecolor{mycolor2}{rgb}{0.00000,0.66667,1.00000}%
\definecolor{mycolor3}{rgb}{0.66667,1.00000,0.00000}%
\definecolor{mycolor4}{rgb}{1.00000,0.66667,0.00000}%
\definecolor{mycolor5}{rgb}{0.66667,0.00000,1.00000}%
\definecolor{mycolor6}{rgb}{1.00000,0.00000,0.66667}%
\definecolor{black}{rgb}{0,0,0}
\begin{tikzpicture}

\begin{axis}[%
width=2.8in,
height=1.3in,
at={(0.758in,0.481in)},
scale only axis,
clip=false,
xmin=0,
xmax=750,
xtick={150,300,450,600, 750},
xlabel style={font=\color{white!15!black}},
xlabel={Mission time in $\si{\second}$},
ymin=0.5,
ymax=3.5,
ytick={1,2,3},
yticklabels={{$r_1$},{$r_2$},{$r_3$}},
ylabel style={at={(0.05,0.5)}, font=\color{white!15!black}},
ylabel={Robots},
axis background/.style={fill=white},
title style={font=\bfseries},
axis x line*=bottom,
axis y line*=left,
legend style={legend cell align=left, align=left, draw=white!15!black}
]
\addplot [color=red]
  table[row sep=crcr]{%
0	1\\
15	1\\
};

\addplot [color=red, line width=4.0pt]
  table[row sep=crcr]{%
15.1012091133719	1\\
115.101209113372	1\\
};

\node[right, align=left, font=\color{gray}]
at (axis cs:-6,1.2) {$\downarrow$};
\node[right, align=left, font=\color{gray}]
at (axis cs:35.101,1.25) {$t^{\mathcal{A}}_1$};
\addplot [color=green]
  table[row sep=crcr]{%
115.101209113372	1\\
142.101209113372	1\\
};

\addplot [color=green, line width=4.0pt]
  table[row sep=crcr]{%
142.353457901472	1\\
242.353457901472	1\\
};

\node[right, align=left, font=\color{gray}]
at (axis cs:121.353,1.2) {$\downarrow$};
\node[right, align=left, font=\color{gray}]
at (axis cs:162.353,1.25) {$t^{\mathcal{B}}_4$};
\addplot [color=mycolor1]
  table[row sep=crcr]{%
242.353457901472	1\\
256.353457901472	1\\
};

\addplot [color=green, line width=4.0pt]
  table[row sep=crcr]{%
256.684760415854	1\\
356.684760415854	1\\
};

\node[right, align=left, font=\color{gray}]
at (axis cs:235.685,1.2) {$\downarrow$};
\node[right, align=left, font=\color{gray}]
at (axis cs:276.685,1.25) {$t^{\mathcal{B}}_5$};
\addplot [color=mycolor2]
  table[row sep=crcr]{%
356.684760415854	1\\
365.684760415854	1\\
};

\addplot [color=mycolor2, line width=4.0pt]
  table[row sep=crcr]{%
365.99212704276	1\\
465.99212704276	1\\
};

\node[right, align=left, font=\color{gray}]
at (axis cs:343.992,1.2) {$\downarrow$};
\node[right, align=left, font=\color{gray}]
at (axis cs:385.992,1.25) {$t^{\mathcal{C}}_6$};
\addplot [color=mycolor2]
  table[row sep=crcr]{%
465.99212704276	1\\
475.99212704276	1\\
};

\addplot [color=black, line width=4.0pt, dashed]
  table[row sep=crcr]{%
476.785475018101	1\\
481.785475018101	1\\
};

\addplot [color=mycolor2, line width=4.0pt]
  table[row sep=crcr]{%
482.182149005772	1\\
582.182149005772	1\\
};

\node[right, align=left, font=\color{gray}]
at (axis cs:460.182,1.2) {$\downarrow$};
\node[right, align=left, font=\color{gray}]
at (axis cs:502.182,1.25) {$t^{\mathcal{C}}_7$};

\addplot [color=red]
  table[row sep=crcr]{%
0	2\\
15	2\\
};

\addplot [color=red, line width=4.0pt]
  table[row sep=crcr]{%
15.6879736715837	2\\
115.687973671584	2\\
};

\node[right, align=left, font=\color{gray}]
at (axis cs:-5.8,2.2) {$\downarrow$};
\node[right, align=left, font=\color{gray}]
at (axis cs:35.688,2.25) {$t^{\mathcal{A}}_3$};
\addplot [color=green]
  table[row sep=crcr]{%
115.687973671584	2\\
126.687973671584	2\\
};

\addplot [color=black, line width=4.0pt, dashed]
  table[row sep=crcr]{%
127.513096849463	2\\
141.513096849463	2\\
};

\addplot [color=green, line width=4.0pt]
  table[row sep=crcr]{%
142.353457901472	2\\
242.353457901472	2\\
};

\node[right, align=left, font=\color{gray}]
at (axis cs:120.353,2.2) {$\downarrow$};
\node[right, align=left, font=\color{gray}]
at (axis cs:162.353,2.25) {$t^{\mathcal{B}}_4$};
\addplot [color=mycolor1]
  table[row sep=crcr]{%
242.353457901472	2\\
256.353457901472	2\\
};

\addplot [color=black, line width=4.0pt, dashed]
  table[row sep=crcr]{%
256.684760415854	2\\
};

\addplot [color=green, line width=4.0pt]
  table[row sep=crcr]{%
256.684760415854	2\\
356.684760415854	2\\
};

\node[right, align=left, font=\color{gray}]
at (axis cs:234.685,2.2) {$\downarrow$};
\node[right, align=left, font=\color{gray}]
at (axis cs:276.685,2.25) {$t^{\mathcal{B}}_5$};
\addplot [color=mycolor4]
  table[row sep=crcr]{%
356.684760415854	2\\
388.684760415854	2\\
};

\addplot [color=black, line width=4.0pt, dashed]
  table[row sep=crcr]{%
388.926329530791	2\\
};

\addplot [color=red, line width=4.0pt]
  table[row sep=crcr]{%
388.926329530791	2\\
488.926329530791	2\\
};

\node[right, align=left, font=\color{gray}]
at (axis cs:366.926,2.2) {$\downarrow$};
\node[right, align=left, font=\color{gray}]
at (axis cs:408.926,2.25) {$t^{\mathcal{A}}_2$};
\addplot [color=mycolor5]
  table[row sep=crcr]{%
488.926329530791	2\\
519.926329530791	2\\
};

\addplot [color=black, line width=4.0pt, dashed]
  table[row sep=crcr]{%
520.29563022754	2\\
600.29563022754	2\\
};

\addplot [color=mycolor5, line width=4.0pt]
  table[row sep=crcr]{%
600.311838585848	2\\
700.311838585848	2\\
};

\node[right, align=left, font=\color{gray}]
at (axis cs:579.312,2.2) {$\downarrow$};
\node[right, align=left, font=\color{gray}]
at (axis cs:620.312,2.25) {$t^{\mathcal{D}}_8$};

\addplot [color=mycolor5]
  table[row sep=crcr]{%
0	3\\
24	3\\
};

\addplot [color=black, line width=4.0pt, dashed]
  table[row sep=crcr]{%
24.5665634747399	3\\
};

\addplot [color=mycolor5, line width=4.0pt]
  table[row sep=crcr]{%
24.56656347474	3\\
224.56656347474	3\\
};

\node[right, align=left, font=\color{gray}]
at (axis cs:3,3.2) {$\downarrow$};
\node[right, align=left, font=\color{gray}]
at (axis cs:95.567,3.25) {$t^{\mathcal{D}}_9$};
\addplot [color=mycolor2]
  table[row sep=crcr]{%
224.56656347474	3\\
266.56656347474	3\\
};

\addplot [color=black,line width=4.0pt,  dashed]
  table[row sep=crcr]{%
266.984788214027	3\\
365.984788214027	3\\
};

\addplot [color=mycolor2, line width=4.0pt]
  table[row sep=crcr]{%
365.99212704276	3\\
465.99212704276	3\\
};

\node[right, align=left, font=\color{gray}]
at (axis cs:343.992,3.2) {$\downarrow$};
\node[right, align=left, font=\color{gray}]
at (axis cs:385.992,3.25) {$t^{\mathcal{C}}_6$};
\addplot [color=mycolor2]
  table[row sep=crcr]{%
465.99212704276	3\\
481.99212704276	3\\
};

\addplot [color=mycolor2, line width=4.0pt]
  table[row sep=crcr]{%
482.182149005772	3\\
582.182149005772	3\\
};

\node[right, align=left, font=\color{gray}]
at (axis cs:460.182,3.2) {$\downarrow$};
\node[right, align=left, font=\color{gray}]
at (axis cs:502.182,3.25) {$t^{\mathcal{C}}_7$};
\addplot [color=mycolor5]
  table[row sep=crcr]{%
582.182149005772	3\\
600.182149005772	3\\
};

\addplot [color=black, line width=4.0pt, dashed]
  table[row sep=crcr]{%
600.311838585848	3\\
};

\addplot [color=mycolor5, line width=4.0pt]
  table[row sep=crcr]{%
600.311838585848	3\\
700.311838585848	3\\
};

\node[right, align=left, font=\color{gray}] at (axis cs:579.312,3.2) {$\downarrow$};
\node[right, align=left, font=\color{gray}] at (axis cs:620.312,3.25) {$t^{\mathcal{D}}_8$};

\end{axis}
\end{tikzpicture}%
	\caption{Gantt chart corresponding to the mission graph depicted in Fig. \ref{fig:paths}. Tasks of the same type are identically colored. Traveling times are represented by thin lines and waiting times as dashed black lines. Gray arrows pointing downwards indicate the starting time of each task.}
	\label{fig:gantt}
\end{figure}
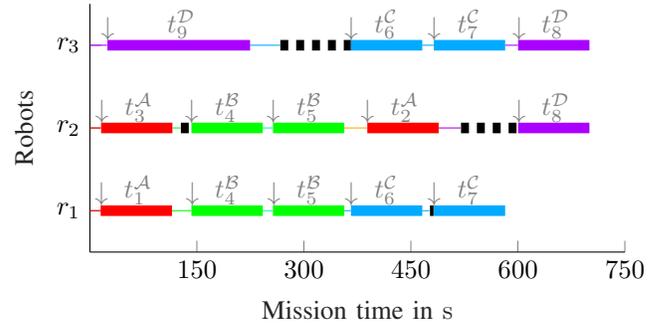

For all tested instances of every problem class, our proposed construction and improvement heuristic yield feasible mission plans. Fig. \ref{fig:paths} shows the resulting mission graph of an optimized solution of an instance of problem class $3\mathcal{A}2\mathcal{B}\mathcal{C}\mathcal{D}$. Additionally to the requirements for assigning cooperative tasks to capable alliances, which are denoted in
Table \ref{tab::static_duration}, this problem class requires  the precedence constraints $(t^\mathcal{A}_1, t^\mathcal{A}_2)_C$, $(t^\mathcal{A}_3, t^\mathcal{B}_4)_C$  and $(t^\mathcal{C}_6, t^\mathcal{D}_8)_C$ to be fulfilled.  The temporal behavior corresponding to the mission graph depicted in Fig.~\ref{fig:paths} is visualized in the Gantt chart in Fig. \ref{fig:gantt}. Herein tasks of the same type are colored identically. Waiting times are marked with dashed black lines, while traveling times are represented by thin lines in the same color as their succeeding task.

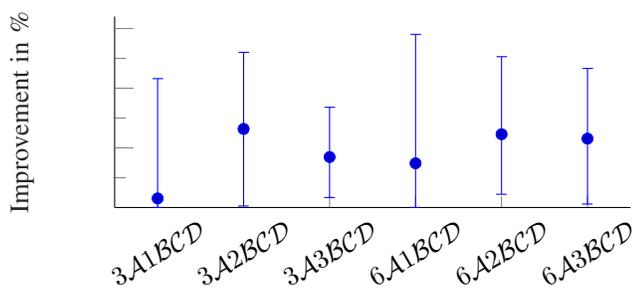
\begin{figure}[tb]
	\centering
%
%

\begin{tikzpicture}

\begin{axis}[%
width=2.7in,
height=1in,
at={(0.758in,0.481in)},
scale only axis,
xmin=0.5,
xmax=6.5,
xticklabels={{},{},{$3\mathcal{A}1\mathcal{B}\mathcal{C}\mathcal{D}$},{$3\mathcal{A}2\mathcal{B}\mathcal{C}\mathcal{D}$},{$3\mathcal{A}3\mathcal{B}\mathcal{C}\mathcal{D}$},{$6\mathcal{A}1\mathcal{B}\mathcal{C}\mathcal{D}$},{$6\mathcal{A}2\mathcal{B}\mathcal{C}\mathcal{D}$},{$6\mathcal{A}3\mathcal{B}\mathcal{C}\mathcal{D}$}},
xlabel style={font=\color{white!15!black}},
ymin=0,
ymax=32,
ytick={5,10,15,20,25,30},
yticklabels = \empty,
extra y ticks = {10,20,30},
extra y tick style = {major tick length = 0.25cm},
ylabel style={font=\color{white!15!black}},
ylabel={Improvement in \%},
axis background/.style={fill=white},
axis x line*=bottom,
axis y line*=left,
legend style={at={(0.5, -0.4)}, anchor = north, legend columns = -1, draw=white!15!black},
x tick label style={rotate=30},
]

\addplot+[forget plot,only marks] plot[error bars/.cd, y dir=plus, y explicit]
coordinates{
	(1,1.52) +- (0,20.08)
	(2,13.17) +- (0,12.84) 
	(3,8.46) +- (0,8.34)
	(4,7.41) +- (0,21.61)
	(5,12.28) +- (0,13.00)
	(6,11.54) +- (0,11.77)
};

\addplot+[forget plot,only marks] plot[error bars/.cd, y dir=minus, y explicit]
coordinates{
	(1,1.52) +- (0,1.52)
	(2,13.17) +- (0,12.92) 
	(3,8.46) +- (0,6.79)
	(4,7.41) +- (0,7.41)
	(5,12.28) +- (0,10.06)
	(6,11.54) +- (0,10.96)
};

\end{axis}
\end{tikzpicture}%
	\caption{Average improvement of the optimized solution compared to the inital solution depicted as blue dots. The minimum and maximum improvement are represented as lower and upper bound of the blue lines.}
	\label{fig:improvement}
\end{figure}

An assessment of the proposed improvement heuristic based on the evaluation of 100 problem instances of each problem class is given in Fig. \ref{fig:improvement}. 
It can be seen that applying the improvement heuristic leads to an average improvement of around \SI{10}{\percent} and a maximum improvement of almost \SI{30}{\percent}. For the smallest problem class $3\mathcal{A}1\mathcal{B}\mathcal{C}\mathcal{D}$ the average improvement drops to approximately \SI{1.5}{\percent}.

The computation times for the developed constructive heuristic and the neighborhood-based improvement heuristic are given in Table~\ref{tab:computationtime}. It can be seen, that the constructive heuristic finds valid solutions in $0.01s$ to $0.05s$. The computational effort for the improvement heuristic increases noticeably with increasing problem sizes. 

\begin{table}[tb]
	\centering
	\begin{tabular}{ccc}
		\hline
		Problem Class & Constr. Heuristic & Impr. Heuristic \\
		\hline
		$3\mathcal{A}1\mathcal{B}\mathcal{C}\mathcal{D}$ & 0.01s & 0.29s \\
		$3\mathcal{A}2\mathcal{B}\mathcal{C}\mathcal{D}$ & 0.02s & 4.10s \\
		$3\mathcal{A}3\mathcal{B}\mathcal{C}\mathcal{D}$ & 0.04s & 19.58s \\
		$6\mathcal{A}1\mathcal{B}\mathcal{C}\mathcal{D}$ & 0.02s & 2.00s \\
		$6\mathcal{A}2\mathcal{B}\mathcal{C}\mathcal{D}$ & 0.04s & 12.04s \\
		$6\mathcal{A}3\mathcal{B}\mathcal{C}\mathcal{D}$ & 0.05s & 36.58s\\
		\hline
	\end{tabular}
	\vspace{0.3cm}
	\caption{Average computation times of construction and improvement heuristic}
	\label{tab:computationtime}
\end{table}

\subsection{Discussion}
The simulation results show that the constructive heuristic yields feasible initial solutions independent of the problem size. The assessment of the improvement heuristic depicted in Fig. \ref{fig:improvement} reveals that applying the improvement heuristic has a high potential to further improve the initial solution especially for larger problem sizes. Nevertheless, there is a big gap between the maximum and the minimum improvement of different instances within a certain problem class. This is most likely due to the fact that some initial solutions are close to a local optimum, while others are not. This in turn influences the improvement that can be achieved by subsequent local search. To further improve our solution approach modifications to the neighborhood operator based on our feasibility criterion and the application of other improvement heuristics, that are able to escape local optima, are conceivable.

\section{Conclusion}\label{sec:conclusion}

In this paper we presented new insights to the feasibility of mission plans for time-extended multi-robot task allocation and scheduling problems with cooperative tasks and precedence constraints. We gave an easy-to-implement criterion to verify the feasibility of mission plans and proposed a constructive and an improvement heuristic working with a generalized objective function structure. We demonstrated the effectiveness of the proposed method by evaluating it using several generalized problem classes of different size. The results show that both the constructive as well as the improvement heuristic yield feasible mission plans and that the local search in average yields significant improvements. In future research we will focus on more neighborhood-operators based on the introduced feasibility criterion and apply more sophisticated improvement heuristics to further improve the results. Furthermore, we will put focus on improving calculation time to allow for solving larger problem instances.

\balance
\bibliographystyle{IEEEtran}
\bibliography{bibliography}

\end{document}